\documentclass[journal,twoside,web]{IEEEtran}
\usepackage{xcolor}
\usepackage{cite}
\usepackage{amsmath,amssymb,amsfonts}
\usepackage{graphicx}
\usepackage{algorithm,algorithmic}
\usepackage{hyperref}
\hypersetup{hidelinks=true}
\usepackage{textcomp}
\usepackage{array}
\usepackage{float}
\usepackage{orcidlink}
\usepackage{siunitx}
\newtheorem{myalg}{Algorithm}
\def\refname{References}

\newcommand{\uf}{\mathbf{u}}

\newcommand{\x}{\textbf{x}}

\newcommand{\nv}{\vec{\boldsymbol{n}}}
\newcommand{\mE}[1]{\mathcal{E}(#1)}
\newcommand{\E}{\mathcal{E}}
\newcommand{\mI}{\mathcal{I}}

\newcommand{\gf}{\mathbf{g}}
\newcommand{\gfD}{\mathbf{g}_D}
\newcommand{\gfT}{\mathbf{g}_T}
\newcommand{\ff}{\mathbf{f}}

\def\BibTeX{{\rm B\kern-.05em{\sc i\kern-.025em b}\kern-.08em
    T\kern-.1667em\lower.7ex\hbox{E}\kern-.125emX}}
\begin{document}

\title{Quantitative Multi-Modal Optical Coherence Photoacoustic Elastography}

\author{Ekaterina Sherina\orcidlink{0000-0002-9542-5145},
        Lisa Krainz\orcidlink{0000-0003-4436-8205},
        Wolfgang Drexler\orcidlink{0000-0002-3557-6398} and Otmar Scherzer\orcidlink{0000-0001-9378-7452}
\thanks{Manuscript received XXX; revised XXX.}
\thanks{The work was supported by the Austrian Science Fund (FWF) in the projects F6801-N36, F6803-N36 and F6807-N36 within the Special Research Programme SFB F68: ``Tomography Across the Scales''. The financial support by the Austrian Federal Ministry for Digital and Economic Affairs, the National Foundation for Research, Technology and Development and the Christian Doppler Research Association is gratefully acknowledged.}
\thanks{Ekaterina Sherina and Otmar Scherzer are with the Faculty of Mathematics, University of Vienna, and Christian Doppler Laboratory for Mathematical Modeling and Simulation of Next Generations of Ultrasound Devices (MaMSi), Oskar Morgenstern-Platz 1, 1090 Vienna, Austria (ekaterina.sherina@univie.ac.at; otmar.scherzer@univie.ac.at). Otmar Scherzer is with the Johann Radon Institute for Computational and Applied Mathematics (RICAM), Altenbergerstrasse 69, A-4040 Linz, Austria (otmar.scherzer@ricam.oeaw.ac.at)}
\thanks{Lisa Krainz and Wolfgang Drexler are with the Center for Medical Physics and Biomedical Engineering, Medical University of Vienna, W\"ahringer G\"urtel, AKH 4L, Vienna, 1090, Austria (lisa.krainz@meduniwien.ac.at; wolfgang.drexler@meduniwien.ac.at)}
\thanks{Ekaterina Sherina and Lisa Krainz are co-first authors. Corresponding author: Ekaterina Sherina (ekaterina.sherina@univie.ac.at)}
\thanks{\textbf{This work has been submitted to the IEEE for possible publication. Copyright may be transferred without notice, after which this version may no longer be accessible.}}
}

\maketitle

\begin{abstract}
We present a novel multi-modal optical coherence photoacoustic elastography framework, which combines two imaging modalities, optical coherence tomography  and photoacoustic tomography, to enable complementary absorption-scattering measurements for the extraction of quantitative tissue features via quasi-static elastography. For this, we develop a sophisticated hybrid inversion algorithm for merging the complementary information layers contained in both optical coherence and photoacoustic-based elastography measurements, and perform systematic evaluations to assess the impact of hybrid elastography data on strain and stiffness reconstructions. An extension to a photoacoustic tomography and optical coherence tomography imaging system allows precise elastographic experiments. Studies on a silicone elastomer phantom demonstrate that the combined optical coherence - photoacoustic approach outperforms single-modality optical coherence elastography and photoacoustic elastography, yielding higher strain signal-to-noise ratios and improved stiffness estimates. These results establish the advantage of multi-modal complementary imaging and  data merging for accurate, high-resolution elastographic strain and stiffness mapping in both scattering and absorbing materials.
\end{abstract}


\section{Introduction}\label{sec:introduction}

\IEEEPARstart{E}{lastography} is a non-invasive imaging approach, which aims at mapping the distribution of mechanical properties within a sample, based on conventional depth-resolved imaging techniques \cite{singh_optical_2025,Manduca_Oliphant_Dresner_Mahowald_Kruse_Amromin_Felmlee_Greenleaf_Ehman_2001,singh_photoacoustic_2019}. For diagnostic accuracy, one is interested in quantitative stiffness maps, rather than qualitative images. Medical elastography has been used in clinical practice for about 25 years. Commercial elastography systems, employing ultrasound (US) or magnetic resonance imaging (MRI), are routinely used for medical diagnosis. Whereas both US and MRI elastography perform well on the scale of whole organs, their resolution is insufficient to resolve micrometer-sized features. Optical coherence elastography (OCE) and photoacoustic elastography (PAE) have been proposed to enable finer-detail stiffness imaging, both showing promising results in clinical research \cite{singh_optical_2025,kennedy_diagnostic_2020, lin_single-breath-hold_2018}. OCE achieves high-resolution stiffness maps at fast imaging speeds, and has been extensively studied in the literature \cite{singh_optical_2025,ZaiMatMatSovHepMowKen21}. It is based on optical coherence tomography (OCT), which uses interferometry to detect backscattered light, producing three-dimensional, depth-resolved morphological images. A wide variety of OCE approaches, differing in loading mechanism, strain detection, and target application, have been proposed \cite{singh_optical_2025}. PAE is based on photoacoustic (PA) imaging (PAI), which records the acoustic response of a sample to absorbed short laser pulses. This work employs photoacoustic tomography (PAT), a variant of PAI with outstanding penetration depth.
Note that whereas the combination of PAI and OCT represents a novel approach in elastography, it has been successfully employed in conventional multi-modal imaging \cite{liu_combined_2016}.

\begin{figure}[!t]
    \centering
    \includegraphics[width=\linewidth]{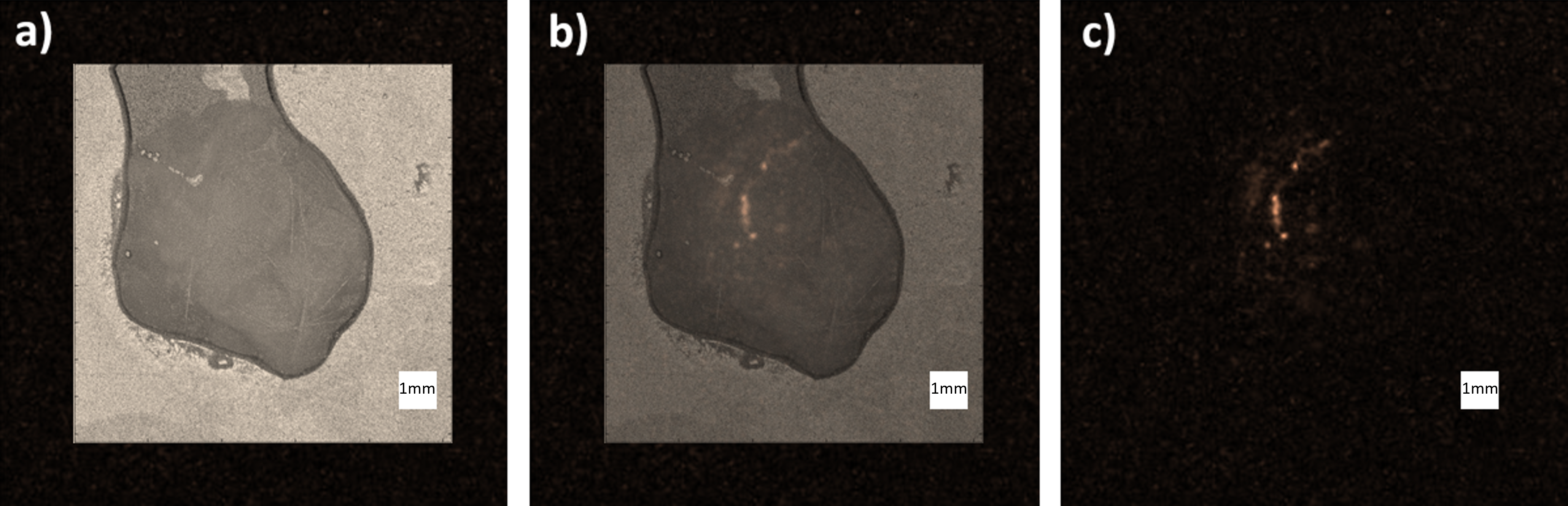}
    \caption{Motivation for multi-modal imaging: Dual‑modal OCT-PAT imaging of a cancer biopsy reveals complementary internal structure, providing information not available from either modality alone. a) OCT; b) OCT and PAT image combined; c) PAT.}
    \label{fig:OCT-PAT-mouse}
\end{figure}

In this work, we present quantitative stiffness and strain reconstruction results from jointly acquired OCT and PAT elastography data. We refer to this imaging modality as \emph{optical coherence photoacoustic elastography} (OCPE). We propose a \emph{multi-layered hybrid inversion algorithm} for OCPE, which leverages dual-modality inputs to quantitatively estimate mechanical properties of the sample. The proposed reconstruction method is adapted to the key features of OCT and PAT by first extracting their complementary information layers from the imaging data and then solving the prior-regularized inverse problem. The hybrid elastography data are fused by the hybrid inversion algorithm based on feature tracking, segmentation, an optical flow (OF) method, and a regularized inverse problem of parameter estimation type.
This work presents an algorithmic and experimental framework for implementing OCPE as an imaging modality and investigates its performance for quantitative imaging. The reconstruction results in Section~\ref{sec:results} demonstrate the advantage of multi-modal OCPE versus single-modality OCE and PAE on phantom studies. OCPE not only enhances the information density available for the stiffness reconstruction, but also addresses the limitations of the individual imaging modalities. When applied separately, OCT is prone to shadow artefacts: highly scattering or absorbing features attenuate the incident light, reducing intensity at depth and producing shadows. PAT is less sensitive to these effects, and can contribute the missing information. PAT visualizes only absorbing features, which are often sparsely distributed in biological samples, e.g., blood vessels in soft tissue. OCT fills these gaps by providing information-rich morphological structure; see Fig.~\ref{fig:OCT-PAT-mouse} for motivation. The quality of the elastographic stiffness reconstruction depends strongly on the information contained in the underlying images, making the combination of the complementary imaging modalities, PAT and OCT, particularly valuable in this context. In addition, our proposed approach is not limited to unidirectional displacement along the imaging axis, as most OCE methods are \cite{singh_optical_2025}, but also accounts for lateral displacement, leading to more precise and comprehensive results, as we have shown previously \cite{Krainz_Sherina_Hubmer_Liu_Drexler_Scherzer_2022}.

\section{Related Work}\label{sec:related} 

To the best of our knowledge, no prior work combines information from more than a single imaging modality to derive elastograms. The terms `dual-modal elastography' and `multi-modal elastography' are commonly used to describe elastography as another information layer in addition to one or more imaging modalities. In such studies, however, only a single imaging modality is used as input for the elastographic reconstruction, which is then added as `another modality'. Although a few publications feature PAT, OCT and elastography, e.g., \cite{yang_learned_2026,yang_photoacoustic_2023}, none of them leverage dual-modality data inputs to estimate mechanical properties of the sample, which constitutes the novelty of this work. The publication \cite{yang_learned_2026} reports to perform multi-modal elastography with OCT and PAM. Both modalities are used for feature learning to differentiate major anatomical regions of the mouse brain. 
However, neither qualitative nor quantitative stiffness maps are provided, raising questions about the classification of \cite{yang_learned_2026} as elastography. In \cite{yang_photoacoustic_2023}, the connection between PA signal and stiffness is leveraged to map the elasticity and viscosity of mouse brain samples via conventional signal analysis. The tissue attenuation is additionally mapped using OCT, yielding a multi-modal technique that relies on a single modality for elastography. In summary, neither of the above publications fully leverage hybrid, complementary OCT and PAT measurements to derive mechanical parameters quantitatively, which is the main contribution of this work.

Quantitative elastography with hybrid data belongs to the class of inverse problems with internal measurements \cite{Kuchment_2012}. Coupling two or more physical modalities allows to indirectly observe interior quantities, which leads to improved uniqueness and stability in reconstructions. In various elastography techniques, imaging modalities provide interior measurements of displacement or strain within the medium \cite{DoyleyMM2012Meas}. Multiple measurements in elastography can further enhance identifiability and stability \cite{Kuchment_2012}. However, in practice,  experiments can be limited to a single uniaxial deformation applied from one direction, unlike the theoretical assumptions. In our experimental setting, we compensate for the lack of additional measurements by extracting the flow field of sample features indicating the general internal motion in the initial steps of our hybrid algorithm. We adapt an OF strategy to incorporate \emph{prior} information for estimating the internal data. The final steps of the proposed hybrid algorithm are based on an inversion of the internal data within a parameter identification inverse problem.

\section{Strain and Stiffness Reconstruction}\label{sec:reconstruction-techniques}
\tikzstyle{process} = [rectangle, draw, text centered, minimum height=3em, align=center, fill=gray!10]
\tikzstyle{processIP} = [rectangle, double, draw, text centered, minimum height=5em, minimum width=5em, align=center, fill=gray!30]
\tikzstyle{connector} = [draw, line width=2pt,->]
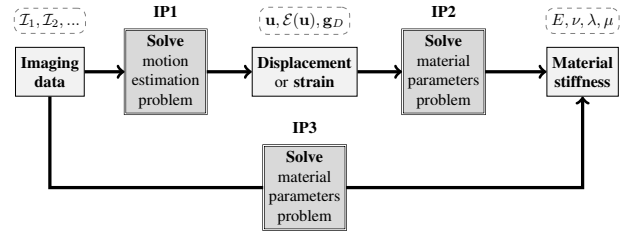
\begin{figure}[!t] \centering
\resizebox{0.45\textwidth}{!}
    {
    \begin{tikzpicture}
        \node [process] at (2.5,-2) (a4) {\textbf{Imaging} \\ \textbf{data}};
        \node [draw=none] at (5,-0.7) {\textbf{IP1}};
        \node [processIP] at (5,-2) (a5) {\textbf{Solve} \\ motion \\ estimation \\ problem};
        \node [process] at (8,-2) (a6) {\textbf{Displacement} \\ or \textbf{strain}};
        \node [draw=none] at (11,-0.7) {\textbf{IP2}};
        \node [processIP] at (11,-2) (a7) {\textbf{Solve} \\ material \\ parameters \\ problem};
        \node [process] at (14,-2) (a8) {\textbf{Material} \\ \textbf{stiffness}};
        \node [processIP] at (8,-4.5) (a9) {\textbf{Solve} \\ material \\ parameters \\ problem};
        \node [draw=none] at (8,-3.2) {\textbf{IP3}};
        \path [connector] (a4) -- (a5);
        \path [connector] (a5) -- (a6);
        \path [connector] (a6) -- (a7);
        \path [connector] (a7) -- (a8);
        \path [connector] (a4) -- (2.5,-4.5) -- (a9) -- (14,-4.5) -- (a8);
        \node [rectangle, draw=gray, rounded corners=5pt, dashed] at (2.5,-0.9) {$\mathcal{I}_1, \mathcal{I}_2,...$};
        \node [rectangle, draw=gray, rounded corners=5pt, dashed] at (8,-0.9) {$\uf, \mE{\uf}, \gf_D$};
        \node [rectangle, draw=gray, rounded corners=5pt, dashed] at (14,-0.9) {$E,\nu,\lambda,\mu$};
    \end{tikzpicture}
    }
    \caption{Schematic depiction of the reconstruction workflow in quasi-static quantitative elastography. Input/output of reconstruction steps, each involving the solution of inverse problems IP1, IP2, IP3 for obtaining material stiffness.}
    \label{fig:1-schematic}
\end{figure}

First, we provide a brief description of the workflow from the perspective of input data and reconstructed quantities. We consider a quasi-static elastography experiment, in which the sample is deformed by an externally applied load. The sample is imaged before and after deformation, following a purposefully designed scanning routine (see Section~\ref{subsec:workflow} for details). The acquired imaging data serve as the primary input to our elastography framework, which aims to quantitatively estimate the \emph{strain} and \emph{stiffness} values within the sample. This can be achieved by following one of two main approaches to quantitative elastography; see Fig.~\ref{fig:1-schematic}. The first and most common one, is known as the \emph{two-step approach}, and involves the consecutive solution of two inverse problems: in inverse problem~1 (IP1), the displacement or strain inside the sample/tissue undergoing deformation is estimated from its imaging data; in inverse problem~2 (IP2), the material parameters of the sample/tissue are then reconstructed from the obtained displacement or strain estimates. The second approach is the so-called \emph{one-step approach} to quantitative elastography: instead of sequentially solving IP1 and IP2, one considers the combined inverse problem~3 (IP3), i.e., the direct reconstruction of the material parameters of the sample/tissue undergoing deformation from its imaging data, without an intermediate calculation of the displacement or strain. Both approaches have their benefits and disadvantages, leading to different reconstruction qualities of stiffness; see Section~\ref{subsec:stiffness-result} for further discussion. 
\begin{figure}[!t] \centering
    \includegraphics[width=0.7\linewidth]{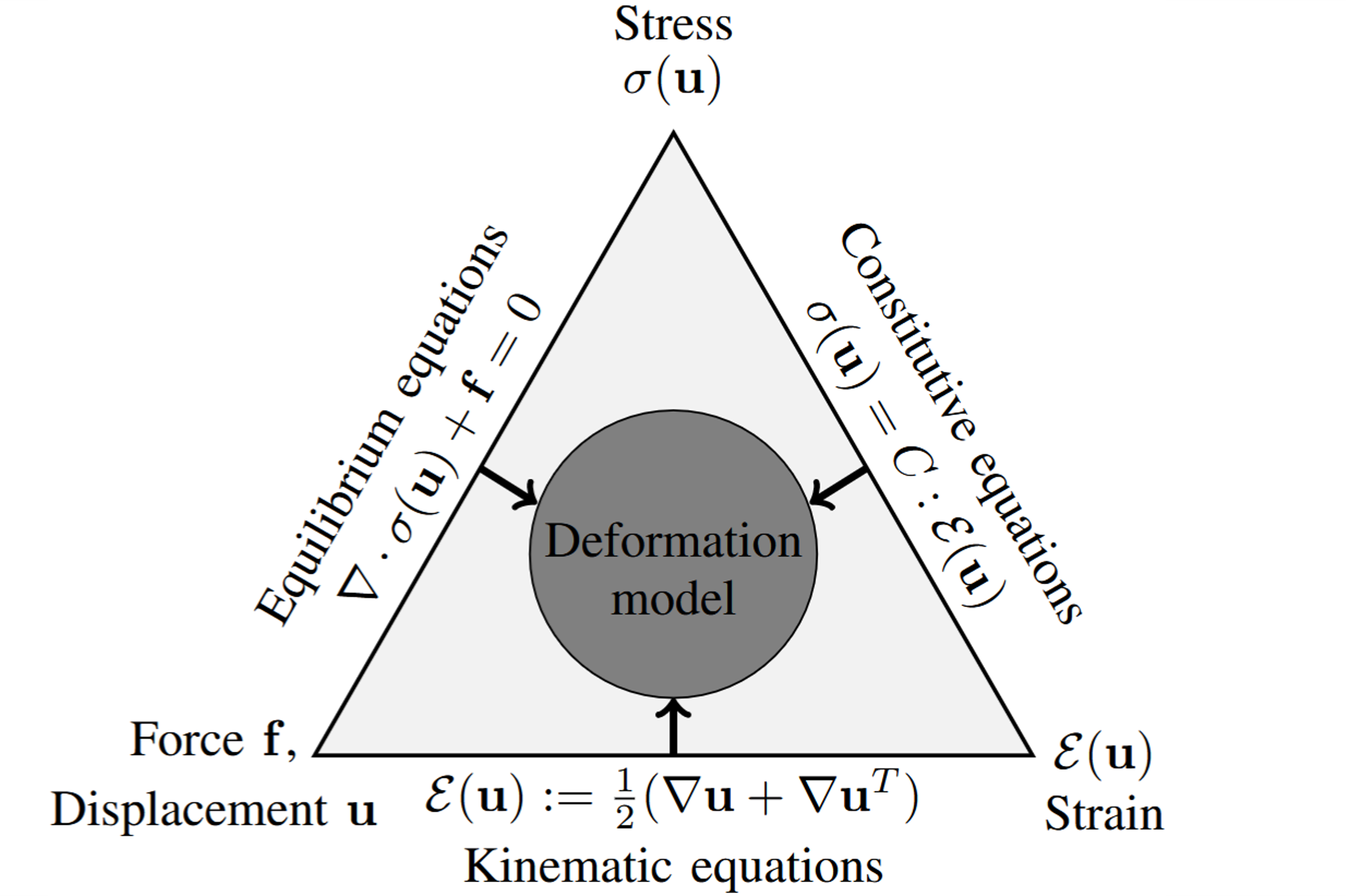}
    \caption{Principle of linear elastography from continuum mechanics.}
    \label{fig:1-cm-principle}
\end{figure}

Furthermore, throughout this paper, we assume that the physical laws governing the deformation experiment follow continuum mechanics and relate several key physical entities (Fig.~\ref{fig:1-cm-principle}): stress and applied load (e.g., force) via the equilibrium equations; stress and strain through the material description via the constitutive equations; and strain and displacement via the kinematic equations. Under the assumptions of infinitesimal strains (i.e., small deformations) and linear isotropic stress-strain relationships, these three equations yield the equations of linearized elasticity \cite{Ciarlet_1994}. More precisely, we assume that the displacement field $\uf$ satisfies the quasi-static linear elasticity equations with displacement-traction boundary conditions, i.e.
\begin{equation}\label{eq:linear-elastic}
    \begin{split}
        -\operatorname{div}{\sigma_{\lambda,\mu}(\uf)} & = \ff \,, \ \text{in} \ \Omega_1 \,, 
        \\ 
        \uf   = \gfD \,, \ \text{on} \ \Gamma_D  \,, \qquad &
        \sigma_{\lambda,\mu}(\uf) \nv = \gfT \,,  \ \text{on} \ \Gamma_T \,.
    \end{split} 
\end{equation}
where $\nv$ is an outward unit normal vector of $\partial\Omega_1 = \overline{\Gamma_D \cup \Gamma_T}$, and $\gfD$ and $\gfT$ are (potentially non-constant) boundary functions defined on $\Gamma_D$ and $\Gamma_T$, respectively. In order to avoid numerical singularities in the FEM solution of \eqref{eq:linear-elastic} at boundary points, where the Dirichlet and Neumann boundary conditions meet, it is beneficial to apply engineering boundary conditions instead. We will expand on this topic in future work. The stress tensor $\sigma_{\lambda,\mu}$ and the strain tensor $\mathcal{E}$ are defined by
\begin{equation} \label{eq:material-law}
    \sigma_{\lambda,\mu}(\uf) := \lambda \operatorname{div}(\uf) \mathbf{I}_n + 2\mu \mathcal{E}(\uf) \,,
\end{equation}
and
\begin{equation}\label{eq:strain}
    \mathcal{E}(\uf) := (\nabla \uf + \nabla \uf^T)/2 \,,
\end{equation}
respectively. Here, $\mathbf{I}_n$ is the identity matrix, and $\lambda, \mu$ are the Lam\'e parameters characterizing the medium. The Lam\'e parameters $\lambda,\mu$ are directly related to the Young's modulus $E$ (i.e.\ stiffness) and the Poison's ratio $\nu$ via
\begin{equation}\label{eq:lame-young-conversion}
    E = \mu(3\lambda + 2\mu)/(\lambda + \mu) \,,
    \quad \text{and} \quad
    \nu = 0.5\lambda/(\lambda + \mu) \,,
\end{equation}
which have physical or diagnostic value, depending on the application \cite{DoyleyMM2012Meas,Manduca_Oliphant_Dresner_Mahowald_Kruse_Amromin_Felmlee_Greenleaf_Ehman_2001}.
Note that the material law \eqref{eq:material-law} is a simplified form of the general Hooke's law
\begin{equation}\label{eq:hookes-law}
    \sigma = C : \mathcal{E} \,, 
\end{equation}
for elastic materials, where $C$ is the fourth-order stiffness tensor. Finally, note that the links between the physical quantities of \eqref{eq:linear-elastic}-\eqref{eq:lame-young-conversion} and the inputs/outputs of IP1, IP2 and IP3 in the reconstruction workflow are indicated in the diagram in Fig.~\ref{fig:1-schematic}, with the corresponding variables highlighted in dashed boxes.

\subsection{Strain}\label{subsec:strain-methods}

In the rapidly evolving field of OCE, numerous methods for computing strain have been developed over the years \cite{singh_optical_2025}. The two prevailing directions of OCE leverage either the phase sensitivity of OCT or conventional cross-correlation techniques of reflectivity-based speckle in the medium \cite{singh_optical_2025}. However, the new multi-modal elastographic OCT-PAT data pose additional challenges for these strain estimation approaches. The PAT acquisition time restricts the deformation experiment to quasi-static loading and comparably large compressions, thus currently excluding the use of phase-sensitive OCT. Moreover, adding absorption contrast to a previously reflectivity-based OCE workflow undermines conventional cross-correlation and requires alternative handling. Consequently, jointly acquired \emph{reflection-absorption data} call for new processing and new displacement and strain estimation techniques. 

In this section, we focus on the estimation of the internal displacement field $\uf$ and strain $\mE{\uf}$ that develop within the sample during the compression experiment, i.e., solving IP1. It is important to note that the reliability of stiffness estimates in two-step approaches depends on the accuracy of the displacement and strain. Hence, we propose a novel \emph{data‑fusing elastographic optical flow method (DEOFM)} that combines OCT and PAT data algorithmically to yield high‑quality displacement and strain maps. This method builds upon our previously developed displacement and strain evaluation technique \cite{Krainz_Sherina_Hubmer_Liu_Drexler_Scherzer_2022,Sherina_Krainz_Hubmer_Drexler_Scherzer_2020} for OCE, whose distinguishing feature is the availability of both lateral and axial strain components, unlike conventional OCE or phase‑sensitive OCE variants. The underlying idea of both approaches is the classic OF equation 
    \begin{equation}\label{eq:OF}
		\mI_t + \nabla \mI \cdot \mathbf{u} = 0 \,, \qquad \text{for all} \quad \, t\ge 0 \,,
	\end{equation}
which connects an image intensity function $\mI = \mI(\mathbf{x},t)$ with a displacement field $\mathbf{u}(\mathbf{x}) = (u_1(\mathbf{x}),u_2(\mathbf{x}))$ for location $\mathbf{x}=(x_1,x_2)$ in $\mathbb{R}^2$ and time $t$ in $\mathbb{R}_+$ in the continuous setting. (In the discrete case, $\mathbf{x}$ stands for the location of an image pixel $i,j$, and $t=1,2,\ldots,m$, indexes images/scans acquired at different times). It is the mathematical foundation of the Horn-Schunck method (see \cite{Szeliski2022}), and we previously proposed its modification into the so-called elastographic optical flow method (EOFM) by incorporating prior knowledge such as: physically meaningful boundary conditions, feature tracking based displacement information, and physically motivated background information. For details, we refer to \cite{Krainz_Sherina_Hubmer_Liu_Drexler_Scherzer_2022}.  

The EOFM approach, when applied to the straightforward fusion of the OCT and PAT scans as $\mI^{\text{OCT+}\zeta\text{PAT}}:= \mI^\text{OCT} +\zeta\mI^\text{PAT}$, is sensitive to the choice of the brightness weight $\zeta$ and to the strategy used for combining (e.g.\ superimposing) the scans. EOFM attempts to fit a single update field to both OCT and PAT images simultaneously, without distinguishing their modality-specific information. The method therefore misfits parts of the estimated displacement field due to the OCT shadowing artefacts and the invisibility of the highly absorbing structures, and to the lack of information about the reflective background medium in PAT. These modality-specific features are not accounted for by EOFM. Hence, we now propose a split-step modification of EOFM to address its shortcomings in application to multi-modal data. We introduce a modality-dependent split step of the update calculation. The proposed DEOFM approach merges OCT and PAT data at the level of the functional formulation, i.e., by fusing both modalities within a single objective functional. First, under a homogeneous medium assumption (as previously), we compute an initial guess $\uf^\text{bg}$, and then search for an update $\uf^\text{upd}$ of the displacement field using OCT data and feature tracking information, thereby reducing the influence of the shadows and highly absorbing structures. We then use the full displacement field from OCT data, obtained by adding the background and the update fields, as an initial background guess for the next step. The new update field is then estimated using PAT data to leverage the more clearly visible absorbing structures of the sample. Finally, the full displacement field $\uf$ is obtained by adding the background $\uf^\text{bg}$ and update $\uf^\text{upd}$ fields. 

Here, we provide a step-by-step outline of the DEOFM algorithm for the displacement estimation by merging the OCT-PAT data of a deformation experiment:
\begin{myalg}[DEOFM] 
\hspace{5pt}
\begin{enumerate}
    \item Let $\mathbf{u}^\text{bg} := \mathbf{u}^\text{hom}$.

    \item Let $k=\text{OCT}$ and solve
    \begin{equation}\label{eq:oflow}
    \begin{split}
        & \min_{\mathbf{u}^\text{upd}} \left( \left\|\mI_t^k + \nabla \mI^k \cdot( \mathbf{u}^\text{upd}
        +  \mathbf{u}^\text{bg})\right\|^2 + \alpha \left\|\nabla \mathbf{u}^\text{upd}\right\|^2 \right.\\
        & +\beta \sum_{i=1}^{n}\int\limits_\Omega g_\sigma(\mathbf{x},\hat{\mathbf{x}}^i)\left\vert (\mathbf{u}^\text{upd} + \mathbf{u}^\text{bg}) - \hat{\mathbf{u}}^i \right\vert ^2 d\mathbf{x} \Big).
    \end{split}
    \end{equation}
    
    \item Calculate $\mathbf{u}^\text{OCT}:= \mathbf{u}^\text{upd} + \mathbf{u}^\text{bg}$.
    
    \item Let $\mathbf{u}^\text{bg} := \mathbf{u}^\text{OCT}$ and solve \eqref{eq:oflow} with $k = \text{PAT}$.

    \item Calculate $\mathbf{u}^\text{PAT}:= \mathbf{u}^\text{upd} + \mathbf{u}^\text{OCT}$.    
    \end{enumerate}
\end{myalg}
The algorithm starts by computing a background displacement field under a homogeneous material assumption. In Step~2, we compute the update field $\uf^\text{upd}$ as the solution of the minimization problem based on the OCT scans and the speckle-tracking vectors $\hat{\uf}^i,i=1,\ldots,n,$ (see Section~\ref{subsec:tracking}) weighted by the regularization parameter $\beta$, under the smoothness assumption adjusted by the regularization parameter $\alpha$. Next, we calculate the full displacement adding the background and update fields and use it as an initial background guess for Step~4. Then, we search for an update field $\uf^\text{upd}$ that minimizes the functional for PAT data, together with the smoothness assumption and the speckle-tracking information. Lastly, the full displacement field from both OCT and PAT data is computed as the sum of the OCT-based background field $\uf^\text{OCT}$ and the PAT-based update field $\uf^\text{upd}$. The DEOFM split-step update reduces interference of information, which occurs in case of using the merged scans, and improves the overall robustness of the displacement estimation method by leveraging strong features of both imaging modalities. Finally, the strain field $\mE{\uf}$ is calculated from the reconstructed displacement field $\uf$ according to \eqref{eq:strain}.

\subsection{Stiffness}\label{subsec:stiffness-methods}
We follow the one-step and two-step approaches to quantitative elastography as described in the introduction to Section~\ref{sec:reconstruction-techniques}. In both cases, we have to solve a so-called inverse problem of parameter estimation type: given observations (data), find the causes of the phenomenon (parameters). In elastography, this means that we are interested in finding the unknown material parameters, given observations of the material behavior, by solving the inverse problems IP2 or IP3, see Fig.~\ref{fig:1-schematic} and Fig.~\ref{fig:1-cm-principle}. Note that both inverse problems are ill-posed  
\cite{BarbonePaulE2002Qeiw,DoyleyMM2012Meas} and thus sensitive to even small amounts of noise in the data. Hence, one must carefully choose methods for obtaining (approximate) solutions to these inverse problems and understand their limitations. We discuss the reconstruction methods for IP2 in Sections~\ref{subsubsec:stiffness-method-direct}, \ref{subsubsec:stiffness-method-nli}, and for IP3 in Section~\ref{subsubsec:stiffness-method-iim}.

\subsubsection{Direct Strain Inversion Method}\label{subsubsec:stiffness-method-direct}

In a quasi-static elastographic experiment, the internal stress distribution is unknown and cannot be measured directly. In case of uniaxial compression, a uniform stress is commonly assumed in order to estimate Young’s modulus from the strain maps \cite{singh_optical_2025}. This assumption ignores the inclusion’s effect on the stress field, as well as any boundary influences. The applied compression force recorded during the experiment is divided by the contact area between the sample and the compression plate to obtain the uniform value of stress $\sigma$. Then, following the general Hooke's law \eqref{eq:hookes-law}, a simplified inversion formula for estimating the Young's modulus $E(\x)$ from the strain map $\E(\x)$ at each point $\x$ (or each pixel, in the discrete case) writes as
\begin{equation}\label{eq:hookes-law-simplified}
    E(\x) = \sigma / \E(\x)\,,
\end{equation}
where the strain $\E(\x)$ is calculated using \eqref{eq:strain} from the displacement field provided by our DEOFM algorithm from Section~\ref{subsec:strain-methods}. This reconstruction approach is called direct strain inversion (DSI).

\subsubsection{Nonlinear Landweber Iteration Method}\label{subsubsec:stiffness-method-nli}

Recall the boundary value problem of linear elasticity \eqref{eq:linear-elastic}, relating the displacement field $\uf$ and the Lam\'e parameters $\lambda$ and $\mu$ at each point $\x$ (or each pixel, in the discrete case). As mentioned above, the inverse problem IP2, see Fig.~\ref{fig:1-schematic}, can be formulated as the problem of solving the nonlinear equation
\begin{equation}\label{eq:ip-operator}
    F(\lambda,\mu) = \uf\,,
\end{equation}
for the unknown Lam\'e parameters $\lambda,\mu$, given (noisy) measurements of the displacement field $\uf^\delta$, where $\delta$ denotes the noise level, meaning that $\|\uf^\delta-\uf\| \leq \delta$, and the operator $F$ represents the model equations 
\eqref{eq:linear-elastic}. The displacement field $\uf^\delta$ is provided by our DEOFM approach from Section~\ref{subsec:strain-methods}. Since IP2 is ill-posed and thus sensitive to noise in the data, we employ nonlinear Landweber iteration (NLI), a well-known iterative regularization methods for solving nonlinear inverse problems \cite{Kaltenbacher_Neubauer_Scherzer_2008}, to compute (approximate) solutions of \eqref{eq:ip-operator}:
\begin{equation}\label{eq:nli}
\begin{split}
    \begin{pmatrix}    
    \lambda_{k+1} \\ \mu_{k+1}
    \end{pmatrix}
    = \begin{pmatrix} \lambda_k \\ \mu_k \end{pmatrix} - \omega^\delta_k F'(\lambda_k,\mu_k)^*(F(\lambda_k,\mu_k)-\uf^\delta)\,,    
\end{split}
\end{equation}
where $k$ is the iteration index and $\omega^\delta_k$ is the steepest descent stepsize \cite{Kaltenbacher_Neubauer_Scherzer_2008}. 
In order to implement the iteration \eqref{eq:nli}, we need explicit expressions for the Fr\'echet derivative $F'(\lambda,\mu)(\cdot,\cdot)$ and its adjoint $F'(\lambda,\mu)^*(\cdot,\cdot)$. For the linear elasticity equations \eqref{eq:linear-elastic}, these have been explicitly derived in \cite{Hubmer_Sherina_Neubauer_Scherzer_2018}. 
Finally, the Young's modulus is calculated from the reconstructed Lam\'e parameters using the conversion relation \eqref{eq:lame-young-conversion}.

\subsubsection{Intensity-based Inversion Method}\label{subsubsec:stiffness-method-iim}

Next, we focus on the inverse problem IP3, see Fig.~\ref{fig:1-schematic}, and a one-step approach called the intensity-based inversion method (IIM) introduced in \cite{Krainz_Sherina_Hubmer_Liu_Drexler_Scherzer_2022}, which computes the material parameters directly from the imaging data. The IIM combines image registration together with a model-based, regularized parameter reconstruction approach. This combination has the advantage of avoiding some approximations and derivative computations typically found in two-step approaches, and results in the IIM being generally more stable to measurement noise. As previously, we consider the sample scanned using a tomographic imaging modality before and after deformation resulting in imaging data $\mI_1$ and $\mI_2$, respectively. Given the displacement $\uf$ satisfying the deformation model \eqref{eq:linear-elastic}, we define the deformation $G(\lambda,\mu)(\x):=\x+\uf(\x,\lambda,\mu)$ at each point $\x$ of the sample. We observe that the material parameters $\lambda,\mu$ are connected to the measured intensities $\mI_1$ and $\mI_2$ via the deformation as
\begin{equation}\label{eq:iim-idea}
    (\mI_2\circ G(\lambda,\mu))(\x) = \mI_2(\x+\uf(\x,\lambda,\mu)) = \mI_1(\x)\,.
\end{equation}
Hence, the IIM is generally formulated as the minimization problem of estimating the unknown parameters $\lambda,\mu$ from 
\begin{equation}\label{eq:iim-min}
    \min_{\lambda,\mu} \, \left( \|\mI_2\circ G(\lambda,\mu) - \mI_1\|^2_{L^2(\Omega)} + \alpha \|(\lambda,\mu)\|^2_{L^2(\Omega)} \right)\,,
\end{equation}
where $\alpha$ is a suitably chosen regularization parameter, controlling the smoothness of  the estimates. For our experiments, we use $\alpha=10^{-3}$. Recall that in the EOFM algorithm, we have used the OF equation, which is the linearization of the image registration equation underlying the IIM method. At first, it may seem counterintuitive to use a more accurate image registration approach for IP3 than for IP1. However, this makes sense, since in IP3 we are registering geometrical objects, which is a difficult task for an OF method, see \cite{Szeliski2022}. 
Next, we leverage the jointly acquired reflection-absorption OCT-PAT data and update the IIM algorithm via data fusion. First, for the image intensities $\mI_1$ and $\mI_2$, we use the combined OCT-PAT scans $\mI^{\text{OCT+}\zeta\text{PAT}}_1$ and $\mI^{\text{OCT+}\zeta\text{PAT}}_2$. Since PAT data contain highly absorbing structures, typically as several visible inclusions embedded in a homogeneous background, we use this additional information to reduce the dimensionality of the problem. To do so, we partition the sample into the inclusion and background areas $D_1,\ldots,D_K$, and estimate the unknown values of $\lambda_k,\mu_k$ for each area $D_k, k=1,\ldots,K$, by describing them as piece-wise constant functions
\begin{equation}\label{eq:parameters-subdomains}
    \lambda(\x) = \sum_{k=1}^K \lambda_k \chi_{D_k}(\x) \quad
    \text{and} \quad \mu(\x) = \sum_{k=1}^K \mu_k \chi_{D_k}(\x)\,,
\end{equation}
where $\chi_{D_k}$ denotes the indicator function of the area $D_k$. Finally, to estimate the Lam\'e parameters $\lambda_k,\mu_k$ corresponding to the inclusions and background areas via the IIM, we solve the following modified minimization problem 
\begin{equation}\label{eq:iim-min-fusion}
\begin{split}
        &\min_{(\lambda_k,\mu_k)^K_{k=1}} \, \left( \Big\|\mI^{\text{OCT+}\zeta\text{PAT}}_2\circ G\left(\sum_{k=1}^K(\lambda_k,\mu_k)\chi_{D_k}\right) \right. \Big. 
        \\ 
        & \left.\Big. - \mI^{\text{OCT+}\zeta\text{PAT}}_1\Big\|^2 + \alpha \Big\|\sum_{k=1}^K(\lambda_k,\mu_k)\chi_{D_k}\Big\|^2 \right)\,.
\end{split}
\end{equation}
For implementation details and a full convergence analysis of the IIM approach, \eqref{eq:iim-min} and \eqref{eq:iim-min-fusion}, we refer to \cite{Krainz_Sherina_Hubmer_Liu_Drexler_Scherzer_2022,SheHub26}. Note that the Young's modulus is calculated from the reconstructed Lam\'e parameters using the conversion relation \eqref{eq:lame-young-conversion}.

\section{Experimental Setup and Workflow}\label{sec:experimental-setup-workflow}
The experimental setup, measurement workflow and data processing pipeline were designed specifically to realize our multi‑modal OCPE technique.

\subsection{Experimental Setup}\label{subsec:experiment}

\begin{figure}[!t]
    \centering
    \includegraphics[width=0.4\textwidth]{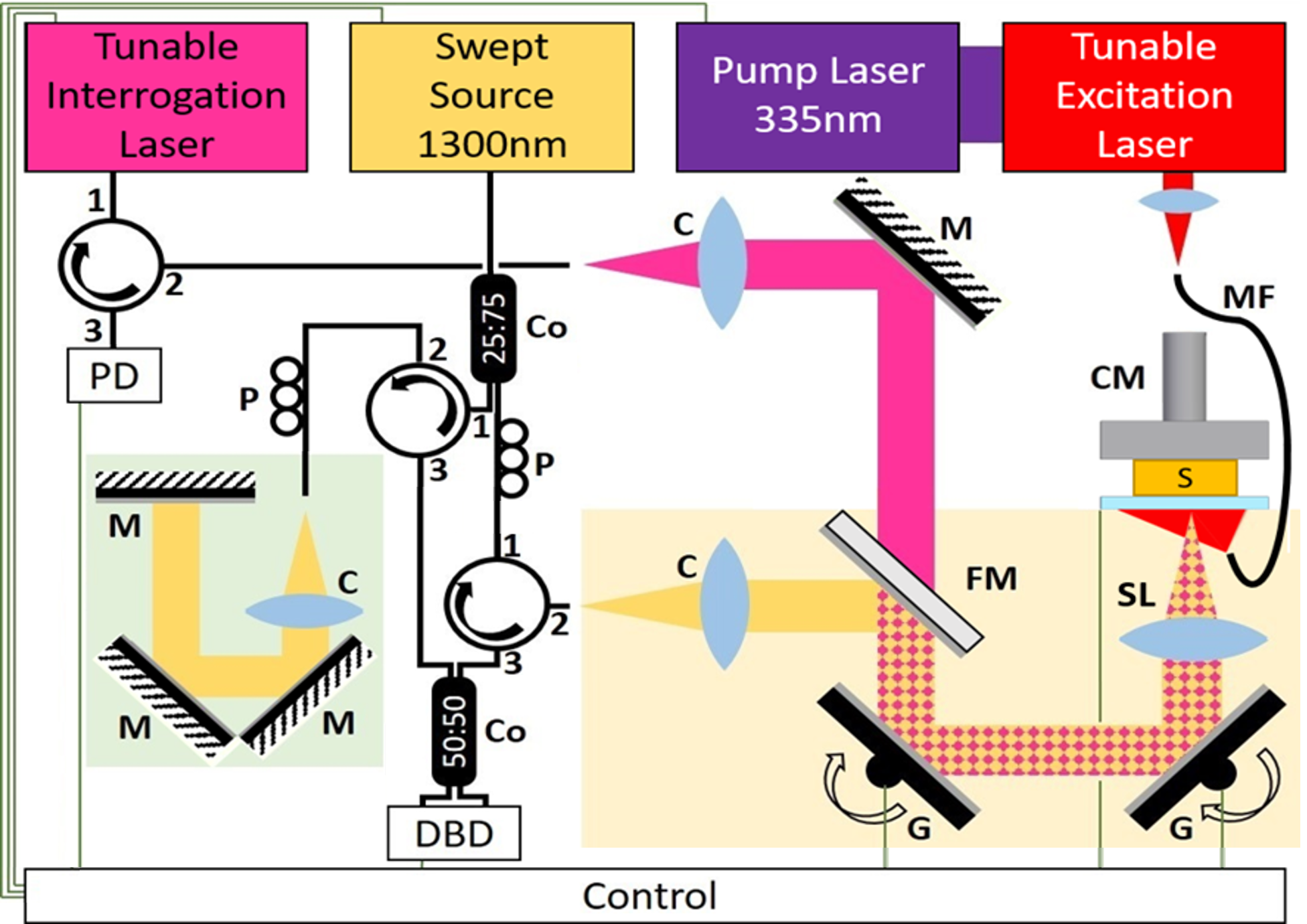}  
    \caption{System Schematics: PAT interrogation recorded by a photodetector (PD) in pink, PAT excitation with multimode fiber (MF) in red, and OCT illumination in yellow. A flip mirror (FM) switches between PAT and OCT. OCT reference arm highlighted in green and sample arm in yellow. Compression module (CM) with sample (S) is explained in detail in Fig.~\ref{fig:compressionhead}. Fiber couplers (Co), polarization paddles (P), dual-balanced detector (DBD), Mirrors (M), collimators (C) and galvanometers (G).}
    \label{fig:system}
\end{figure}
\begin{figure}[!t]
    \centering
    \includegraphics[width=0.4\textwidth]{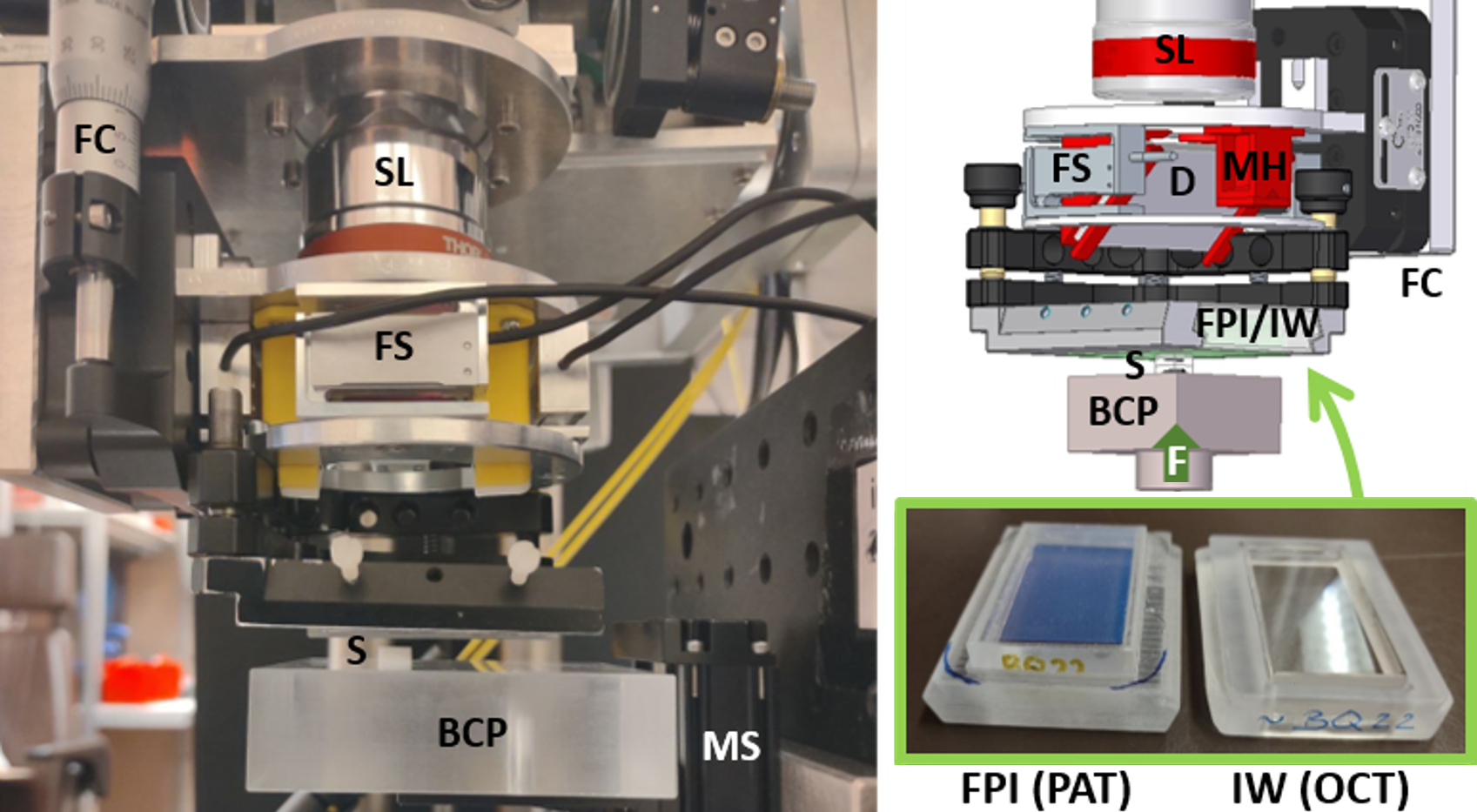}  
    \caption{OCPE Imaging Head: A specifically designed compression module; Scan lens (SL) for both OCT illumination and PAT interrogation; Three force sensors (FS); Sample (S); Motorized stage (MS); Back compression plate (BCP). Switching of Fabry-P\'{e}rot interferometer (FPI) for PAT and imaging window (IW) for OCT via pull-out mechanism\cite{liu_combined_2016}; Multi-modal fiber holder (MH) for PAT excitation; Dichroic mirror (D); Focus control (FC).}\label{fig:compressionhead}
\end{figure}
A combined PAT-OCT device\cite{chen_non-invasive_2017} was upgraded to enable elastographic measurements (see Fig.~\ref{fig:system}). The OCT part uses an akinetic swept-source (SLE-101, Insight Photonic Solutions, US) with a central wavelength of \SI{1300}{nm} and a bandwidth of \SI{30}{nm} and a sweep repetition rate of \SI{222}{kHz}.
The fiber-based system operates in a Mach-Zehnder configuration. The combined signal is recorded by a dual-balanced detector (BPD-1, Insight Photonic Solutions, US) and digitized by a data acquisition card (ATS9360, Alazar Technologies Inc., Canada). The OCT part of the system achieves a lateral resolution of \SI{24}{\micro m} and an axial resolution of \SI{27}{\micro m} in biological tissue. 
The PAT part uses a \SI{100}{Hz} pulsed laser  (SpitLight 600 OPO, INNOLAS, Germany) operated at a wavelength of \SI{808}{nm} used to excite the sample. The sample's acoustic response is detected by a Fabry-P\'{e}rot interferometric ultrasound sensor (FPI), provided by Paul C. Beard from the University College London (UCL, London, UK) \cite{liu_combined_2016}. This signal is read out optically by an interrogation laser (Tunics T100S-HP-CL, Yenista Optics, France) and recorded by a photodiode (G9801-22, Hamamatsu Photonics K.K., Japan) connected to a custom-built transimpedance amplifier and digitized by a data acquisition card (NI PCI 5114, National Instruments, US). The PAT part reaches a lateral resolution of \SI{62}{\micro m} and an axial resolution of \SI{31}{\micro m}.
The OCT free-space sample arm shares its beam path with the PAT interrogation laser, ensuring a similar field of view (FoV) and a motorized flip mirror (Model 8892, New Focus, US) is used to switch between modalities. The FPI is mounted via a sliding mechanism, allowing easy replacement by an imaging window for OCT.

The imaging probe head is equipped with three force sensors (KD34s $\pm$10 N, ME-Systeme, Germany). In this configuration, the compression force is picked up on the same side from which the images are taken (see Fig.~\ref{fig:compressionhead}). This allows to assess not only small phantoms and biopsies, but also large and \emph{in vivo} samples. The mechanical loading for small sample elastography is applied via a motorized stage (T-LSM050 A, Zaber Technologies, Canada), which is synchronized with PAT and OCT image acquisition.
All experiments are controlled using LabView (2018, National Instruments, US) and tomographic image reconstruction for PAT and OCT post-processing is performed in MATLAB (R2021a, Mathworks, US). 

\subsection{Measurement Workflow}\label{subsec:workflow}

\begin{table}[!t]
    \caption{Total applied strain between compression step 1 and consecutive compression steps 2-10, in \%}
    \label{tab:dp-strains}
    \centering
    \begin{tabular}{c|c|c|c|c|c|c|c|c|c}
        \textbf{cs} & 2 & 3 & 4 & 5 & \textbf{6} & 7 & 8 & 9 & 10 \\
        \hline
        \textbf{1} & 1.05 & 1.75 & 2.78 & 3.85 & \textbf{4.55} & 5.24 & 6.26 & 6.99 & 7.69
    \end{tabular}
\end{table}
Instead of the displacement-controlled loading used in our previous work \cite{Krainz_Sherina_Hubmer_Liu_Drexler_Scherzer_2022,Sherina_Krainz_Hubmer_Drexler_Scherzer_2020}, a force-controlled loading scheme was implemented to ensure that the sample undergoes the same compression during PAT and OCT imaging. 
First, the motorized stage compresses the sample against the PAT FPI with a force of \SI{100}{mN}. After a waiting time of \SI{3}{minutes}, which ensures that the sample is in mechanical equilibrium to satisfy the quasi-static condition, a three-dimensional PAT image is acquired. 
This process is repeated ten times, with the loads being consecutively increased by \SI{25}{mN}, up to a maximum of \SI{325}{mN}. The resulting strain between compression step 1 and consecutive compression steps 2-10 is shown in Table~\ref{tab:dp-strains}. After acquiring the PAT compression series, the sample is allowed to elastically deform back to its original form. Meanwhile, the FPI is exchanged with the OCT imaging window, and then the same loading-imaging scheme is repeated with OCT. A total of two sets of images are acquired: one set of OCT images and one set of PAT images, visualizing the same ten consecutive compression steps in 3D.

After examining the total applied strain (Table~\ref{tab:dp-strains}), we present our strain and stiffness results in Section~\ref{sec:results} for the deformation between compression steps 1 (cs1) and 6 (cs6), which fall within the linear elastic regime. The elastic approximation is commonly employed for rubber-like materials, for small deformations with  principal strains $\lesssim 5\%$ \cite{Sadd2014}.

\subsection{Phantoms}\label{subsec:phantoms}

\begin{figure}[!t]
    \centering    
    \includegraphics[width=0.49\textwidth]{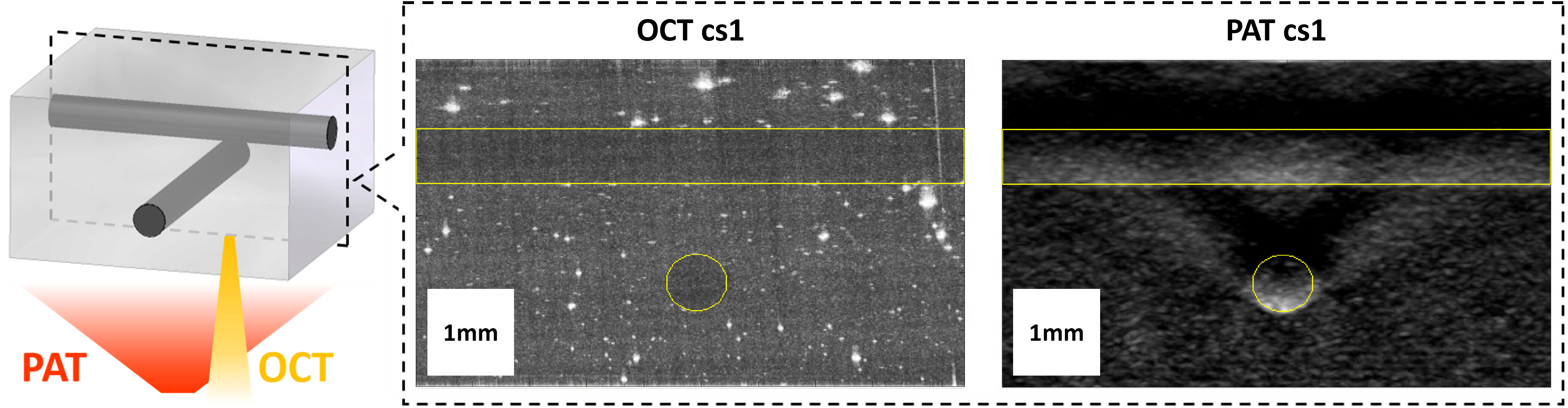} 
    \caption{Phantom and Segmentation: Location of the inclusions, imaging direction, FoV and result of the PAT-based image segmentation (yellow lines), see Section~\ref{subsec:processing} for details.}
    \label{fig:phantom}
\end{figure}
A silicone elastomer phantom was produced specifically for this measurement. It features a soft background material (Eco10, Smooth-On, TX, US) containing titanium dioxide particles (37262-25G, Sigma-Aldrich, CA, US) to mimic the scattering contrast of soft biological tissue, and contains two perpendicularly offset inclusions, representing blood vessels. The inclusions are made of another, harder silicone elastomer (S10, Smooth-On, TX, US) with five drops of black silicone rubber pigment (Shenzhenshi Baishifuyou Trading Co., Ltd., Shenzhen, China) added per \SI{40}{g} of silicone, making the inclusions slightly absorbing in PAT. The ground truth Young's modulus values were determined using a tensile testing machine consisting of a 200N ElectroForce TestBench system (TA Instruments, New Castle, DE, US) and a 10N load cell (Model WMCP-1000 g, Interface Inc., AZ, US).

\subsection{Data Processing}\label{subsec:processing}

OCT and PAT images are reconstructed from their respective raw datasets using standard reconstruction algorithms. The OCT image is rotated to ensure that the top and bottom surfaces of the sample are orthogonal to the image's depth axis. Afterwards, PAT and OCT pixel sizes are matched. OCT images inherently exhibit an image intensity decline with depth, which can be a challenge for motion estimations based on OF. This problem is mitigated by applying a depth-dependent intensity compensation. We restrict the analysis to a region of interest (ROI) within the phantom, to exclude boundary artefacts. To allow OCT-PAT co-registration, fiducial markings were applied on the back compression plate with a permanent marker. The absorbing inclusions are prominently visible in PAT and faintly visible in OCT, thereby adding additional information for the rigid body transformation that is used to align the PAT and OCT images. Finally, the PAT images are cropped laterally to the same ROI as the OCT images. 
Then, in axial direction, data outside the phantom region above the glass window and the back compression plate are removed.

For DEOFM analysis, we generate a 2D maximum intensity projection (MIP) from the 3D ROI for both PAT and OCT by selecting central slices and leveraging the experiment's symmetry to map the cylindrical inclusions to a circle and a bar. This step is necessary, since there is an inherent ambiguity in 3D motion determination via OF. We cross-check symmetry by analysing the 3D sparse displacement fields from feature tracking (see Section~\ref{subsec:tracking}), searching for  symmetry planes in the data. The MIPs high and low intensity bounds are rescaled so that the most frequent intensities align across all compression steps after the projection. As noted in Section~\ref{subsec:strain-methods}, co-registered OCT-PAT reflection and absorption data require distinct processing to leverage their dual nature. We further process the PAT images by first denoising them to suppress background noise. The PAT images are then segmented into inclusions and background regions (see Fig.~\ref{fig:phantom}). When selecting automated segmentation algorithms, we leverage \emph{a priori} structural sample information. For more general inclusions, shape-agnostic segmentation algorithms are required \cite{PhamDzungL2000}. When image quality is unreliable, manual marking can serve as a baseline. For the segmentation of our PAT data, we produce an average intensity projection over the several selected slices of the 3D PAT ROI and use MATLAB's Image Segmenter (MATLAB\copyright Toolstrip: Image Processing and Computer Vision): the Find Circles algorithm is applied for the circular inclusion, and Active Contours is used for the bar inclusion.  The segmentation masks are used to remove the remaining image reconstruction artefacts from the PAT images. As mentioned in Section~\ref{subsec:strain-methods}, we examined different ways of combining co-registered OCT and PAT images while developing our proposed displacement/strain estimation DEOFM approach. In our data-processing tests, the optimal choice of the weighting coefficient is $\zeta=0.65$, which provides the best trade-off between OCT contrast and PAT structural details.

\subsection{Feature Tracking}\label{subsec:tracking}
\begin{figure}[!t]
    \centering    
    \includegraphics[width=0.6\linewidth]{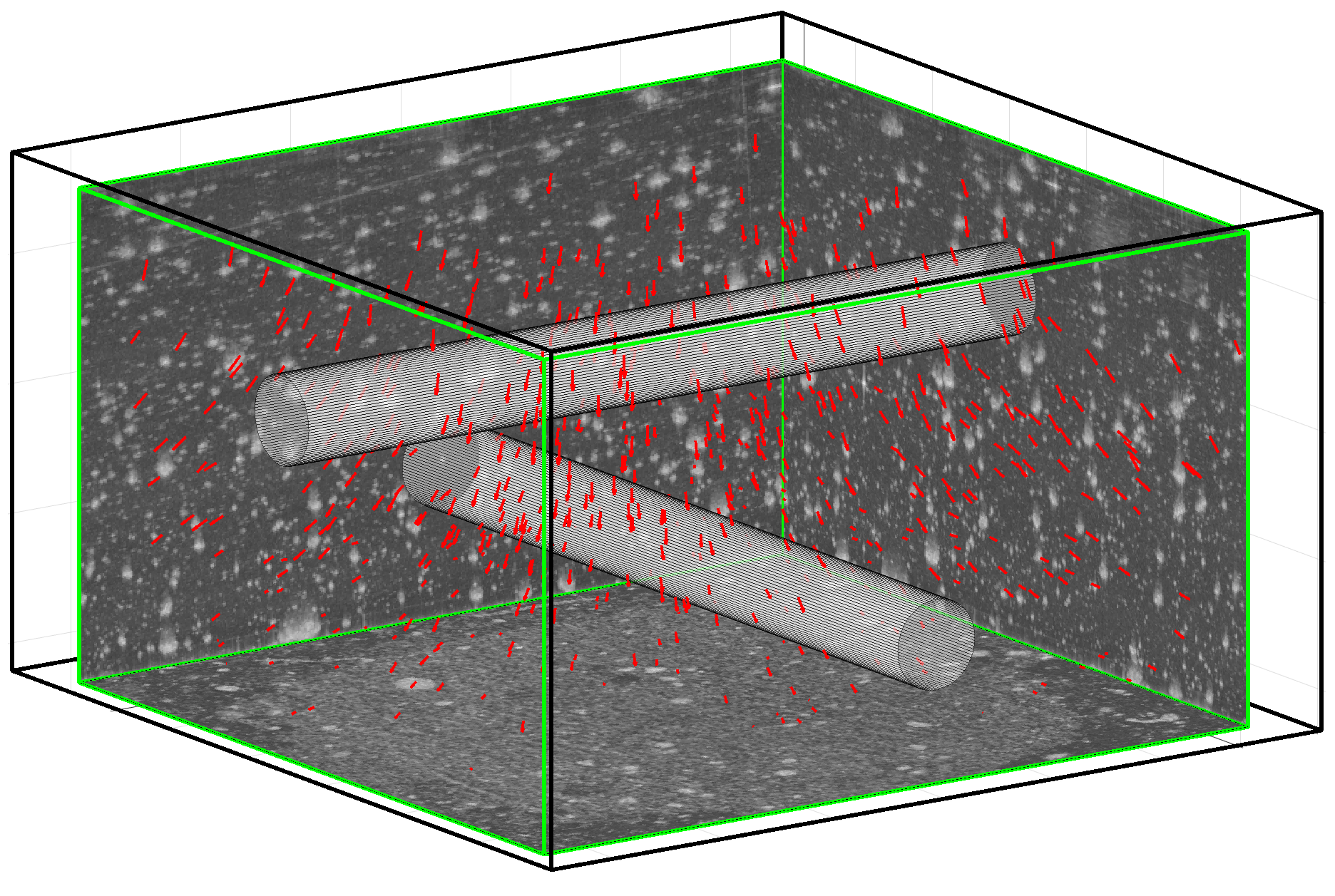}
    \resizebox{\linewidth}{!}{
    \begin{tabular}{c|c|c|c}
        \hline
        Modality & Compression & No.\ features & No.\ displacement vectors \\
        \hline
        OCE & cs1-cs6 & 1548, 1795 & 580 \\
        \hline
    \end{tabular}}
    \caption{Feature tracking between compression steps 1 (cs1) and 6 (cs6). Top: Tracked titanium dioxide scatterers movement, depicted as displacement vectors (red). Bottom: Tracking statistics: total numbers of detected features in cs1 and cs6 and total identified displacement vectors for cs1-cs6.}
    \label{fig:dp-tracking}
\end{figure}
For DEOFM analysis in Section~\ref{subsec:strain-methods}, we require \emph{a priori} information about the general internal motion in the sample inferred from titanium dioxide scatterers, i.e., the additional regularization term involving the vectors $\hat{\uf}^i$ at the locations $\hat{\x}^i$. For this, we use the heuristic-based image-processing algorithm for feature tracking proposed in \cite{Sherina_Krainz_Hubmer_Drexler_Scherzer_2020}, which is purposefully designed to detect and track titanium dioxide scatterers. We apply the algorithm to pairs of successive 3D OCT volumes. Fig.~\ref{fig:dp-tracking} illustrates the detected scatterers and their tracked movement, i.e., the displacement vectors, between compression step 1 and 6. For implementation details, we refer to \cite{Sherina_Krainz_Hubmer_Drexler_Scherzer_2020}.

\section{Results \& Discussion}\label{sec:results}
The capability of our novel, multi-modal OCPE framework (see Section~\ref{sec:experimental-setup-workflow}) together with the reconstruction techniques introduced in Section~\ref{sec:reconstruction-techniques} is demonstrated for the deformation between the compression steps 1 (cs1) and 6 (cs6), which fall within the linear elastic regime (see Section~\ref{subsec:workflow}).

\subsection{Strain}\label{subsec:strain-result}

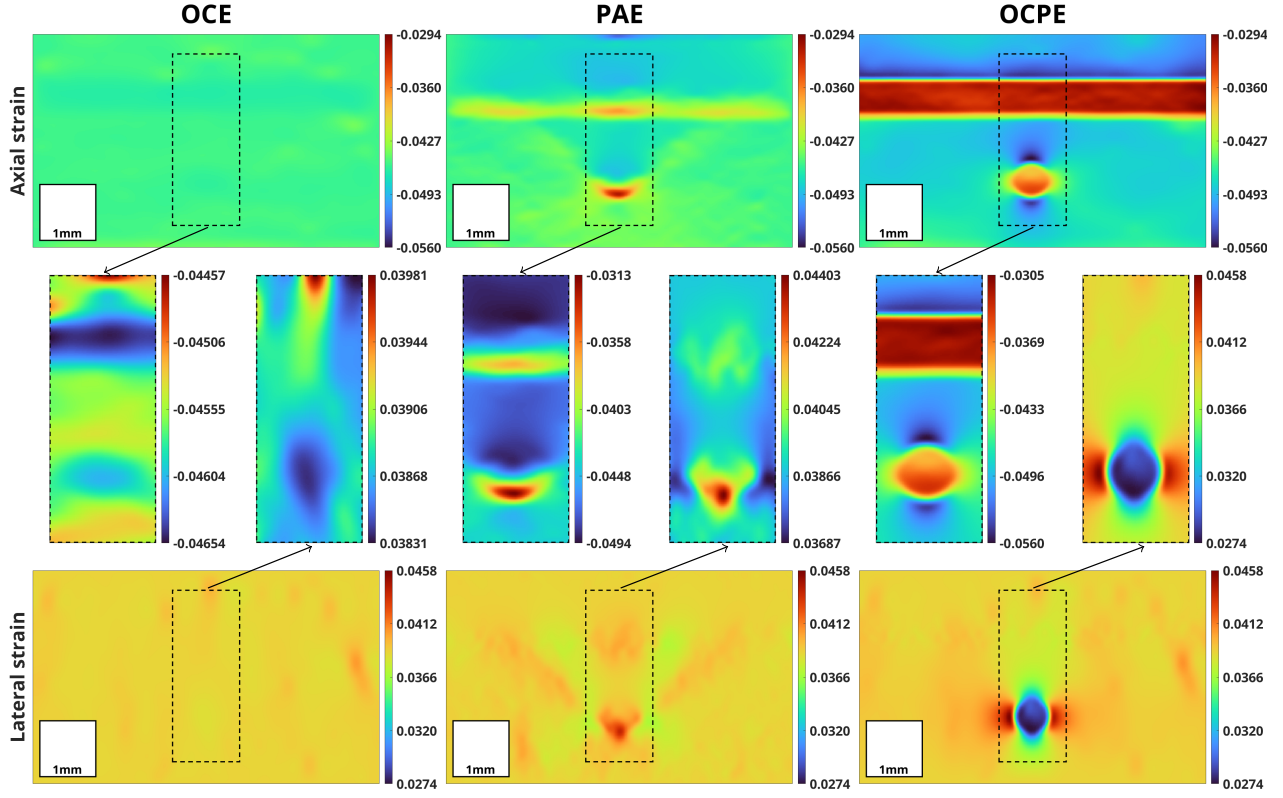
\begin{figure*}[!t]
    \centering    
    \resizebox{0.93\textwidth}{!}{
    \begin{tikzpicture}
        \node [inner sep=0pt] at (0pt,0pt) (a1) {\includegraphics[width=\textwidth, clip=true, trim={40pt 160pt 20pt 60pt}]{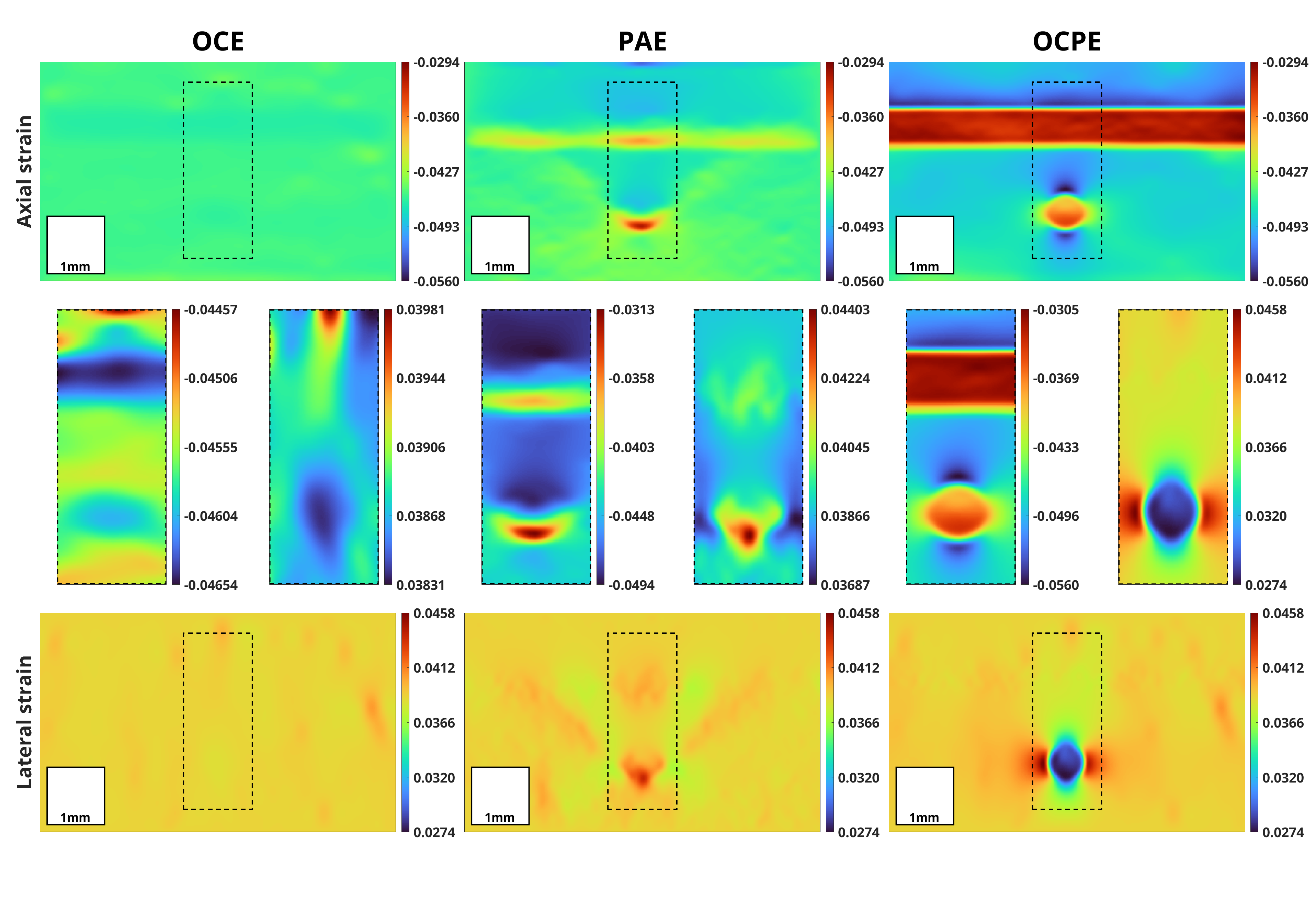}};
        \path[draw, line width=0.5pt,->] (-6.2,2.4) -- (-7.7,1.75);
        \path[draw, line width=0.5pt,->] (-0.25,2.4) -- (-1.7,1.75);
        \path[draw, line width=0.5pt,->] (5.7,2.4) -- (4.3,1.75);
        \path[draw, line width=0.5pt,->] (-6.2,-2.8) -- (-4.7,-2.2);
        \path[draw, line width=0.5pt,->] (-0.25,-2.8) -- (1.3,-2.2);
        \path[draw, line width=0.5pt,->] (5.7,-2.8) -- (7.3,-2.2);
    \end{tikzpicture}
}
    \caption{Axial and lateral strain reconstruction: OCE captures background motion but is compromised by shadow artefacts from the inclusions (left). PAE detects inclusion motion yet lacks high information density (middle). OCPE, combinging OCT and PAT imaging for elastography, recovers the expected strain distribution within the sample (right).}
    \label{fig:strain-results-axial-lateral}
\end{figure*}

First, we reconstruct the strain from OCPE imaging data via DEOFM (Section~\ref{subsec:strain-methods}) and compare it with strain maps from single-modality OCE and PAE estimated via EOFM. All reconstructions were performed in Python on a Linux workstation (24 \texttimes 12th Gen Intel\textregistered Core\texttrademark i9-12900K, 62.6 GiB of RAM) with an average wall‑clock runtime of 3–5 minutes per reconstruction. The resulting axial and lateral strains are visualized in Fig.~\ref{fig:strain-results-axial-lateral}, and already clearly show the benefits of our novel multi-modal elastography approach. 
OCE detects the regions of lower axial strain within the phantom, but strongly overestimates the inclusion strain.  PAE is limited to the information available from PAT. The PAT signal is strongest on the side of the absorbing inclusion that is closest to the excitation (see Fig.~\ref{fig:phantom}). This intensity gradient, together with the missing background information, results in an incomplete mapping of the inclusion strain. The zoom region insets in Fig.~\ref{fig:strain-results-axial-lateral} illustrate the strain maps with an adaptive color scale, indicating that the inclusions are indeed visible in OCE and PAE. Benefiting from higher information density, OCPE is capable of reconstructing a more accurate and realistic strain distribution. 
Across all results in Fig.~\ref{fig:strain-results-axial-lateral}, we observe that the inclusion oriented parallel to the imaging plane is not visible in the lateral strain maps. We attribute this effect to the stiff inclusion acting as lateral constraint that inhibits lateral deformation in the surrounding material. The stiffness ratio of the background to the inclusion silicone rubber was measured to be approximately 1:6.9 (see Table~\ref{tab:young-modulus-results}), which is considered a significant mechanical contrast. 

Both OCT and PAT can exhibit modality-specific imaging artefacts. In particular, strong absorbers or scatterers reduce the light intensity available for imaging in deeper regions of the sample, leading to a `stripy' appearance of the OCT image. In the most severe case, known as shadow artefacts, the deeper regions are completely dark, preventing the acquisition of morphological information in these areas. PAT can exhibit similar shadow artefacts, but they are generally much less pronounced. In FPI-based PAT, more severe artefacts arise from the PA image reconstruction. The FPI is flat, positioned on one side of the sample, and detects only the part of the sound waves perpendicular to its surface. The reconstruction artefacts originating from this lack of information are negligible in an \textit{en face} view of the 3D data. In a transverse view, which is important for elastographic analysis, the artefacts become more pronounced and appear as `wing-shaped' features (see Fig.~\ref{fig:phantom}). Taken together, these artefacts pose challenges for motion reconstruction techniques based on OF.  
Combining two imaging modalities, as demonstrated with DEOFM, improves the reconstruction quality by fusing complementary image information and mitigating the influence of modality‑specific artefacts, leading to improved strain maps.

\subsection{Stiffness}\label{subsec:stiffness-result}

\begin{table}[!t]
    \caption{Young's modulus reconstructions: mean values (in kPa) over regions defined by PAT segmentation masks}
    \label{tab:young-modulus-results}
    \centering
    \begin{tabular}{c|c|c|c}
        & \textbf{Background} & \textbf{Inclusion1} & \textbf{Inclusion2} \\
        \hline
        Ground truth & 72.3\textpm6.8 & 501.3\textpm14.8 & 501.3\textpm14.8 \\
        DSI reconstruction & 52.5\textpm2.6 & 66.5\textpm6.0 & 77.7\textpm6.0 \\
        NLI reconstruction & 160.3\textpm56.2 & 220.2\textpm17.2 & 367.2\textpm34.4 \\
        \textbf{IIM reconstruction} & \textbf{112.8}\textpm12.4 & \textbf{235.7}\textpm21.9 & \textbf{461.5}\textpm48.0 \\
        \hline
    \end{tabular}
\end{table}

\begin{figure}[!t]
    \centering
    \includegraphics[width=0.9\linewidth]{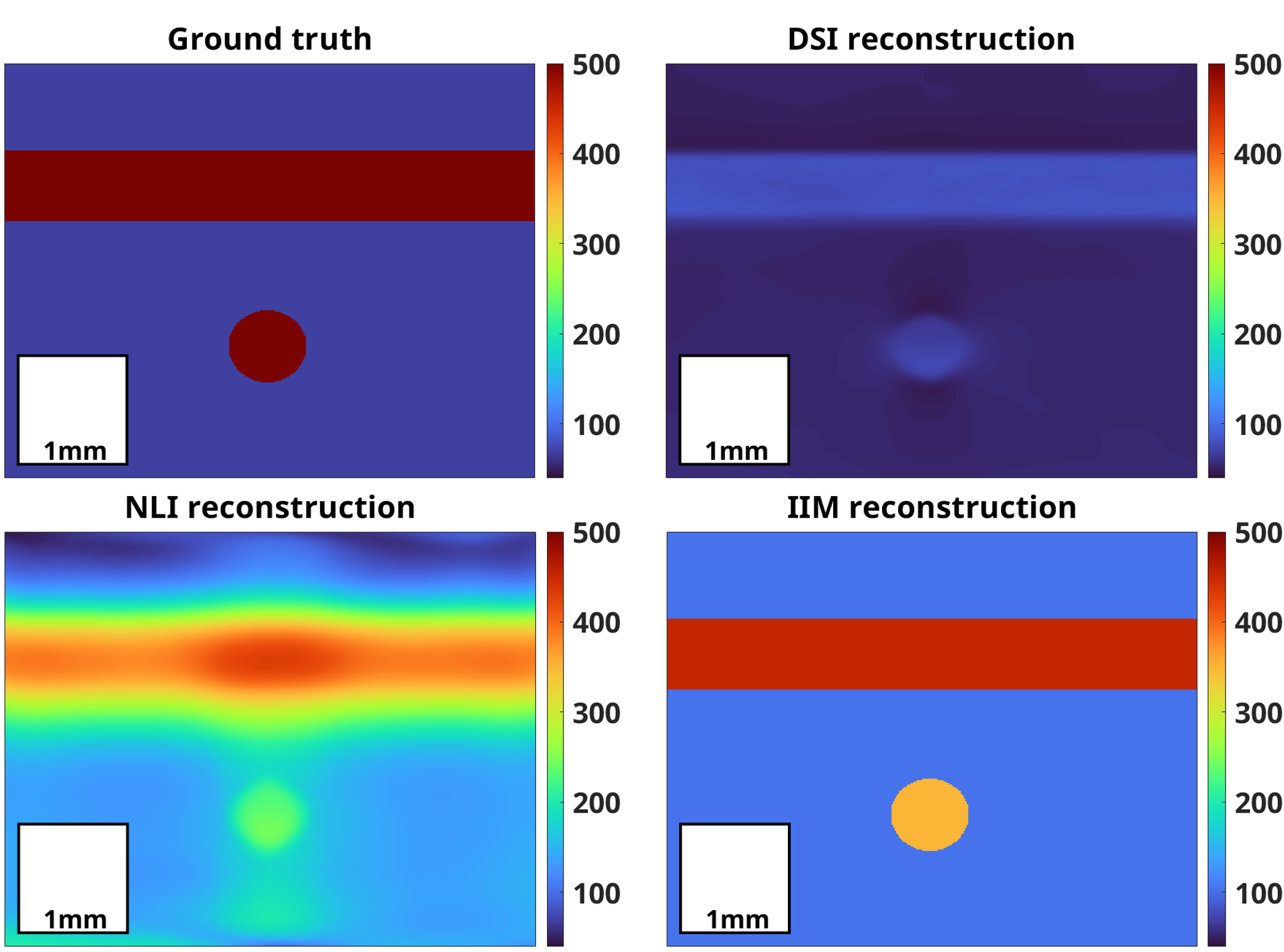}
    \caption{Young's modulus maps (kPa): OCPE stiffness reconstruction results versus ground truth in the phantom obtained using the inversion techniques described in Section~\ref{sec:reconstruction-techniques}; direct strain inversion (DSI); nonlinear Landweber iteration (NLI); intensity-based inversion method (IIM).}
    \label{fig:stiffness-results}
\end{figure}

From OCPE strain and imaging data, we reconstruct the sample's stiffness using the three different inversion techniques described in Section~\ref{subsec:stiffness-methods}: direct strain inversion (DSI), nonlinear Landweber iteration (NLI), and the intensity-based inversion method (IIM). Fig.~\ref{fig:stiffness-results} illustrates the reconstruction results and compares them with the ground truth. DSI and NLI, both spatially dependent methods, accurately localize the inclusions within the sample but yield different stiffness estimates. 
IIM, which solves a reduced-dimensionality minimization problem, relies on PAT segmentation and derives stiffness estimates at specified inclusion locations. The obtained stiffness values for the inclusions and background are summarized in Table~\ref{tab:young-modulus-results}, where the mean values are calculated over regions defined by the PAT segmentation masks (Section~\ref{subsec:processing}) for the DSI and NLI reconstructions. Table~\ref{tab:young-modulus-results} indicates that the resulting stiffness depends on the chosen inversion method and, implicitly, on the input data. Although the two orthogonal inclusions, prepared simultaneously from the same silicone elastomer, should have identical stiffness (based on tensile tests in Section~\ref{subsec:phantoms}), we consistently observe that the stiffness of inclusion 1 (a circle at the bottom) is underestimated by all inversion methods, whereas that of inclusion 2 (a bar at the top) is reconstructed more accurately. In Section~\ref{subsec:projection-direction}, we discuss how stiffness evaluation is influenced by the orientation of the imaging data, the size of the inclusion, and its position relative to the compression and imaging planes.

DSI performs worst overall, underestimating the background stiffness by \SI{19.8}{kPa} and the inclusion stiffness by \SI{429.2}{kPa} on average. As described in Section~\ref{subsec:stiffness-methods}, DSI assumes a constant stress distribution within the sample. This assumption holds only if the stiffness is homogeneous and the boundary of the sample can be neglected. Neither is satisfied in our deformation experiment. The impact of the stress inhomogeneity caused by the inclusions is greatest in the inclusion region, which explains why the inclusion stiffness is determined inaccurately, whereas the background stiffness could be calculated rather precisely. In general, the homogeneous stress assumption should be used with caution in elastography, since it implies that stiffness inhomogeneities are quantified while simultaneously assuming their absence in the data.

The two-step NLI is able to accurately recover the structure of the sample. However, it struggles to retrieve the correct contrast: NLI overestimates the background stiffness by \SI{88.0}{kPa} and underestimates the stiffness of inclusion 1 by \SI{281.1}{kPa} and inclusion 2 by \SI{134.1}{kPa}. 
The NLI approach does not require a constant stress assumption and the relevant information is expected to be encoded in the strain (Fig~\ref{fig:strain-results-axial-lateral}). Hence, the reconstruction quality of NLI depends directly on the strain maps used as the input of IP2. Since the strain first has to be estimated from the sample images by solving IP1, the quality of this estimate depends on the imaging data and the chosen strain evaluation method. Mathematically, IP2 (as well as IP3) require more than one displacement or strain field to ensure a globally unique and high-quality reconstruction of spatially varying stiffness; see \cite{BarbonePaulE2002Qeiw} for details.
Since our experiment is limited to a single uniaxial deformation, we compensate for the lack of additional measurements by using feature tracking information. To mitigate this limitation and improve stiffness estimates, we rely on dimensionality reduction of the problem and avoid computing derivatives from the noisy data (i.e., the calculation of strain from the displacement field in IP2). For this, we solve IP3 by applying the IIM approach directly to the OCT-PAT imaging data and leverage the previously obtained PAT segmentation. Fig.~\ref{fig:stiffness-results} and Table~\ref{tab:young-modulus-results} illustrate that IIM performs best among all reconstruction methods: IIM overestimates the background stiffness by \SI{40.5}{kPa} and underestimates the stiffness of inclusion 1 by \SI{265.6}{kPa} and inclusion 2 by only \SI{39.8}{kPa}. 

For a fair comparison, we also applied all three inversion techniques to the OCT and PAT datasets separately. Both strain-based methods (DSI and NLI) fail to produce reliable stiffness reconstructions, due to a lack of structural features and the limited quality of the OCE and PAE strains. Image-based IIM is robust only when the necessary priors are available. We observed that OCT data without PAT-based segmentation cannot reliably solve the OCE minimization problem since the problem is severely ill-posed. Similarly, PAE alone suffers from a lack of OCT structural information; even when PAT segmentation is available, the problem remains ill-posed. In both OCE and PAE, the local minimizers lie outside of the admissible parameter set (i.e., realistic stiffness constraints). Conversely, the complementary OCT-PAT data contain the necessary prior information for the robust implementation of quantitative elastography. These results demonstrate the advantage of our multi-modal OCPE approach, owing to complementary imaging and data merging.

\subsection{Sample Orientation}\label{subsec:projection-direction}

\begin{figure}[!t]
\centering
    \includegraphics[width=0.49\linewidth]{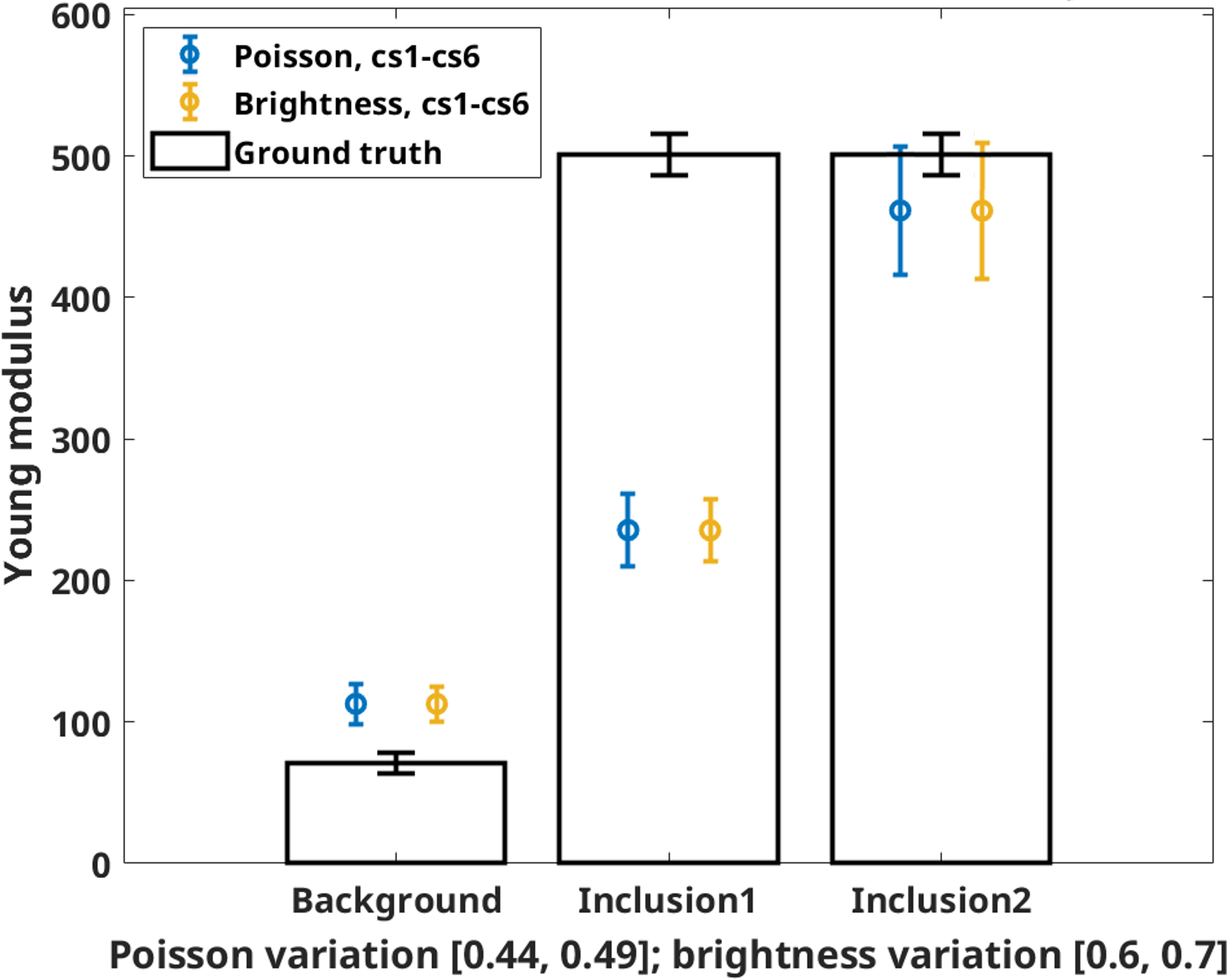}
    \includegraphics[width=0.49\linewidth]{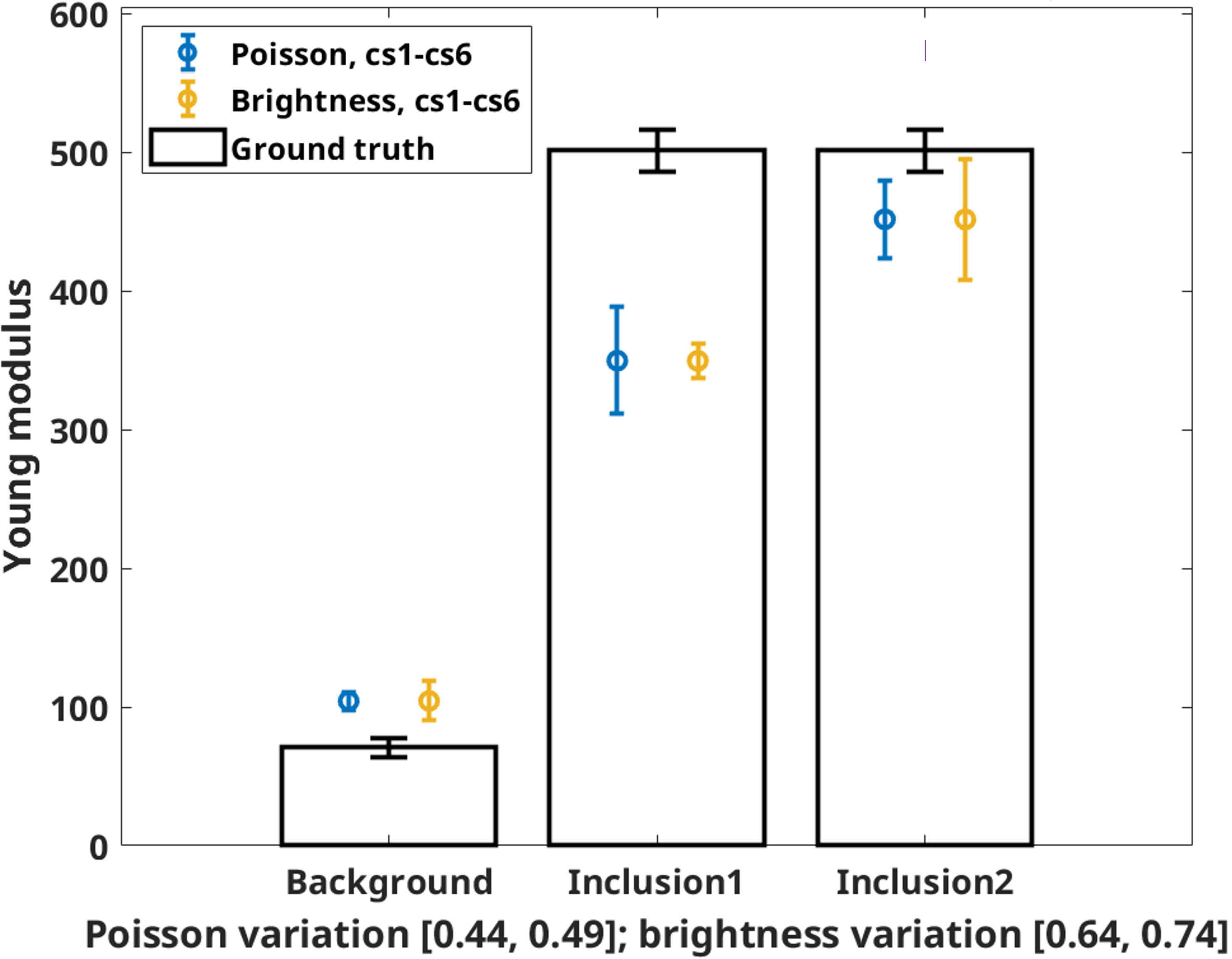}
    \caption{Young's modulus reconstruction results with the IIM for compression experiments cs1-cs6 on MIPs along x (left) and y (right) cross-sections of the phantom. The estimated values for background, inclusion 1 and 2 are compared to the ground truth. Uncertainty is quantified with respect to  Poisson's ratio and OCT-PAT image brightness variations.}
    \label{fig:results_young_iim}
\end{figure}
As mentioned above, we consistently observed that all inversion methods underestimate the stiffness of inclusion 1, although it should have the same stiffness as inclusion 2; see Table~\ref{tab:young-modulus-results}. We infer that stiffness estimates are influenced by the orientation of the imaging data, the inclusion size and its position relative to the compression and imaging planes. To investigate this, we performed several reconstruction tests using IIM on two orientations, taking 2D MIPs along the x- and y-axis of the 3D OCT data volume, respectively. As seen in Fig~\ref{fig:phantom}, the inclusion brightness is depth dependent relative to its location to the imaging plane. Furthermore, the inclusion is subject to different displacement when it is near or far from the compression plate. This has important implications for inverse problems: since the object is probed via an applied force, small or negligible loading (i.e., in the regions far from the compression plate) reduces the observability of stiffness variations and hinders the performance of inversion methods. Note also that the stiffness estimation inverse problem (IP2) requires more than one linearly independent displacement or strain field to ensure a globally unique, high-quality reconstruction. In addition, the problem unavoidably accumulates various errors in the data, such as discretization errors, 3D to 2D projection errors and post-processing error. Taken together, these factors (inclusion area, location, exposure to deformation, image intensity) may affect the stiffness estimates. To support our assumptions, we performed uncertainty quantification for the two test settings with respect to variations in the Poisson's ratio and OCT-PAT image brightness. The variation step size was set to $0.005$ for both parameters. The estimated uncertainties and the distribution of IIM solutions are displayed in two charts in Fig.~\ref{fig:results_young_iim}. The results show that the background stiffness is only slightly affected by parameter variations or by the choice of orientation. We observe a significant improvement in the stiffness estimates for inclusion~1, when the IIM input data are switched to the y-orientation. We attribute this to the larger area of inclusion 1. The stiffness estimates for inclusion 2 remain consistently within the same range across all tests. 

Finally, another contributing factor may be the structure of our phantom: two stiff beam-like, perpendicular inclusions are embedded in a soft background with an approximate stiffness ratio of 1:6.9 (see Table~\ref{tab:young-modulus-results}). These stiff inclusions may act as longitudinal (along the inclusion) constraints, inhibiting transverse deformation in the surrounding material and influencing the deformation of both inclusions \cite{Buryachenko2007}. In particular, inclusion~2, which is closest to the compression plane, may strongly inhibit transverse deformation and influence the reconstructed Young's modulus value of inclusion 1.

\section{Conclusion}\label{sec:conclusions}

In this paper, we introduced a \textit{novel multi-modal optical coherence photoacoustic elastography} approach, termed OCPE, which for the first time combines both OCT and PAT measurements for the extraction of quantitative mechanical parameter maps via quasi-static elastography. We described the OCPE framework and presented our 
multi-layered hybrid inversion algorithm to estimate the strain and to quantify the Young's modulus within a sample. We investigated OCT-PAT data-fusion techniques, and assessed their impact on the quality of strain and stiffness reconstructions. Furthermore, we demonstrated that the OCPE approach outperforms single-modality OCE and PAE on phantom studies, see in particular Fig.~\ref{fig:strain-results-axial-lateral}.
Reconstructions were validated against the ground truth values determined with a commercial device, see Table~\ref{tab:young-modulus-results}. These results establish the advantage of multi-modal absorption-scattering OCT-PAT imaging and data merging for robust, quantitative elastography.

\section*{Acknowledgments}
The authors thank Simon Hubmer (JKU Linz, Austria) for fruitful discussions, Paul Beard and Edward Zhang (UCL, UK) for providing the Fabry-P\'{e}rot interferometer, and Martin Stoiber (MUW, Austria) for assistance with the Bose device.

\bibliographystyle{IEEEtran}
\bibliography{IEEEabrv,references}

\end{document}